%% file: main.tex
\documentclass[preprint,12pt]{elsarticle}

\usepackage{amssymb}
\usepackage{amsmath}
\usepackage{hyperref}
\usepackage{xurl}
\usepackage{pdflscape}
\usepackage{geometry}
\usepackage{array}
\usepackage{setspace}
\usepackage[linesnumbered,ruled,vlined]{algorithm2e}
\usepackage{verbatim}
\usepackage{booktabs}
\usepackage{siunitx} % for proper units

\begin{document}
\renewcommand{\labelenumii}{\arabic{enumi}.\arabic{enumii}}

\begin{frontmatter}

% double check number of words
%TC:ignore
% \section*{Word Counts}
% \detailtexcount{main}
%TC:endignore

%% Title, authors and addresses

%% use the tnoteref command within \title for footnotes;
%% use the tnotetext command for theassociated footnote;
%% use the fnref command within \author or \affiliation for footnotes;
%% use the fntext command for theassociated footnote;
%% use the corref command within \author for corresponding author footnotes;
%% use the cortext command for theassociated footnote;
%% use the ead command for the email address,
%% and the form \ead[url] for the home page:
%% \title{Title\tnoteref{label1}}
%% \tnotetext[label1]{}
%% \author{Name\corref{cor1}\fnref{label2}}
%% \ead{email address}
%% \ead[url]{home page}
%% \fntext[label2]{}
%% \cortext[cor1]{}
%% \affiliation{organization={},
%%             addressline={},
%%             city={},
%%             postcode={},
%%             state={},
%%             country={}}
%% \fntext[label3]{}

\title{Spatial Disparities in Fire Shelter Accessibility: Capacity Challenges in the Palisades and Eaton Fires}

%% use optional labels to link authors explicitly to addresses:
%% \author[label1,label2]{}
%% \affiliation[label1]{organization={},
%%             addressline={},
%%             city={},
%%             postcode={},
%%             state={},
%%             country={}}
%%
%% \affiliation[label2]{organization={},
%%             addressline={},
%%             city={},
%%             postcode={},
%%             state={},
%%             country={}}

\input{sections/authors}

% not exceed 250 words
%% Abstract
\begin{abstract}
The increasing frequency and severity of wildfire in California, exacerbated by prolonged drought and environmental changes, pose significant challenges to urban community resilience and equitable emergency response. The study investigates issues of accessibility to shelters during the Palisades and Eaton Fires which started in January 2025 in Southern California that led to over 180,000 displacements and the loss of 16,000 structures. Despite coordinated efforts of many organizations’ emergency assistance, shelter shortages left many evacuees without safety or accessible refuge. This research aims to measure shelter accessibility during the fires’ peak, evaluate whether existing shelter capacity met the demand, and identify spatial disparities in access. Findings reveal severe shelter shortages and pronounced inequities in access to shelters, particularly in geographically isolated regions and mountainous areas. To address these challenges, we implemented shelter placement strategies using both capacity-based and distance-based approaches, demonstrating potential improvements in accessibility and equity. The findings underscore the critical need for strategic shelter planning and infrastructure development to enhance disaster readiness and reduce vulnerability in regions that frequently experience wildfires.
\end{abstract}

%%Research highlights
% \begin{highlights}
% \item Research highlight 1
% \item Research highlight 2
% \end{highlights}

%% Keywords
\begin{keyword}
%% keywords here, in the form: keyword \sep keyword
spatial accessibility \sep wildfire \sep disparities \sep shelter 

%% PACS codes here, in the form: \PACS code \sep code

%% MSC codes here, in the form: \MSC code \sep code
%% or \MSC[2008] code \sep code (2000 is the default)

\end{keyword}

\end{frontmatter}

%% Add \usepackage{lineno} before \begin{document} and uncomment 
%% following line to enable line numbers
%% \linenumbers

%% main text
%%

%% Use \section commands to start a section
\section{Introduction}
\label{sec:intro}
\input{sections/introduction}

\section{Related Work}
\label{sec:related}
\input{sections/related_work}

\section{Study Area and Data}
\label{sec:study_area}

\input{sections/study_area}

% \section{Data}
% \label{sec:data}
\input{sections/data}

\section{Method}
\label{sec:method}
\input{sections/method}

\section{Results}
\label{sec:results}
\input{sections/results}

\section{Conclusion}
\label{sec:conclusion}
\input{sections/conclusion}

%% Use \subsubsection, \paragraph, \subparagraph commands to 

%% The Appendices part is started with the command \appendix;
%% appendix sections are then done as normal sections
\appendix
% \label{sec:supplementary}
% Supplementary materials are provided in a separate document.
\input{sections/append}

%% If you have bib database file and want bibtex to generate the
%% bibitems, please use
%%
\bibliographystyle{elsarticle-num} 
\bibliography{main}

\end{document}

%% file: sections/authors.tex
%% use the tnoteref command within \title for footnotes;
%% use the tnotetext command for theassociated footnote;
%% use the fnref command within \author or \affiliation for footnotes;
%% use the fntext command for theassociated footnote;
%% use the corref command within \author for corresponding author footnotes;
%% use the cortext command for theassociated footnote;
%% use the ead command for the email address,
%% and the form \ead[url] for the home page:
%% \title{Title\tnoteref{label1}}
%% \tnotetext[label1]{}
%% \author{Name\corref{cor1}\fnref{label2}}
%% \ead{email address}
%% \ead[url]{home page}
%% \fntext[label2]{}
%% \cortext[cor1]{}
%% \affiliation{organization={},
%%             addressline={},
%%             city={},
%%             postcode={},
%%             state={},
%%             country={}}
%% \fntext[label3]{}

\author[txst]{Su Yeon Han} %% Author name
\author[txst]{Yubin Lee} %% Author name

%% Author affiliation
\affiliation[txst]{organization={Department of Geography and Environmental Studies, Texas State University},
            addressline={601 University Drive}, 
            city={San Marcos},
            postcode={78666}, 
            state={TX},
            country={USA}}

\author[emory]{Jooyoung Yoo}
            
\author[kh]{Jeon-Young Kang}
\author[kh]{Jinwoo Park}

\affiliation[kh]{organization={Department of Geography, Kyung Hee University},
            addressline={26 Kyungheedae-ro, Dongdaemun-gu}, 
            city={Seoul},
            postcode={02447}, 
            % state={},
            country={Republic of Korea}}

\author[txst2]{Soe W. Myint}
\author[txst3]{Eunsang Cho}
\author[txst]{Xin Gu}

\affiliation[txst2]{organization={The Meadows Center for Water and the Environment, Texas State University},
            addressline={601 University Drive}, 
            city={San Marcos},
            postcode={78666}, 
            state={TX},
            country={USA}}
            
\affiliation[txst3]{organization={Ingram School of Engineering, Texas State University},
            addressline={601 University Drive}, 
            city={San Marcos},
            postcode={78666}, 
            state={TX},
            country={USA}}

\author[emory]{Joon-Seok Kim\corref{cor1}} %% Author name
\ead{joonseok.kim@emory.edu}
\cortext[cor1]{Corresponding author}
%% Author affiliation
\affiliation[emory]{organization={Department of Computer Science, Emory University},
            addressline={400 Dowman Drive}, 
            city={Atlanta},
            postcode={30322}, 
            state={GA},
            country={USA}}

%% file: sections/introduction.tex
The western United States has experienced more frequent wildfires in recent years with California particularly at risk due to prolonged drought conditions and increased temperatures \cite{cova2013mapping, NASA_WildfireTrends, NOAA_WildfireClimate, williams2019observed}. The growing intensity of wildfires has resulted in disastrous consequences by claiming lives, destroying property, and damaging essential public infrastructure including roads, power, and water systems. \cite{Statista_WildfireDeaths, NBCNews_CaliforniaWildfires, JEC_WildfireCosts, wang2021economic}. On January 7, 2025, the Palisades and Eaton Fires began in Southern California. The Palisades Fire in Los Angeles spread to a size of 23,700 acres by January 12 \cite{calfire_palisades_fire_2025}, while the Eaton Fire near Pasadena expanded to over 14,000 acres within several days \cite{nypost2025}. The Palisades and Eaton Fires continued for 24 days until they were completely contained by January 31, 2025 \cite{NBCNews_PalisadesEatonFire}. The extensive burning period of the Palisades and Eaton Fires led to widespread devastation—the Palisades Fire destroyed 7,000 structures and resulted in 12 fatalities, while the Eaton Fire crushed more than 9,000 structures and claimed 17 lives \cite{lacounty2025fires}.

The Palisades and Eaton wildfires forced a large number of residents to evacuate, leading to overcrowded shelters and leaving many without immediate refuge. \textit{The New York Times} reported that many of them struggled to find shelters during their evacuations \cite{NYTimes2025}. As part of the response efforts, aid organizations such as the American Red Cross established eight emergency shelters throughout Los Angeles County to help with wildfire evacuations \cite{NYTimes2025}. However, the available facilities were insufficient to accommodate the massive demand, which resulted in many evacuees  seeking accommodations from relatives and friends or sleeping in their vehicles \cite{NYTimes2025}. Others attempted to find hotel rooms or short-term rentals but faced difficulties due to a drastic surge in rental demand caused by the massive influx of displaced people \cite{NYTimes2025}. These prolonged shelter shortages forced them to relocate multiple times during the crisis \cite{nyt2025lafires}. Access to shelters was particularly limited in geographically isolated areas and settlements on undulating terrains. Specifically, restricted road access and steep terrain were major factors that delayed prompt evacuations \cite{nytimes2025californiafire, bbc2025evacuation}. These geographic barriers were especially evident in communities such as the Santa Monica Mountains, where narrow roads restricted both evacuation routes and emergency response operations \cite{apnews2025wildfires}.

The immense destruction and chaos caused by the Palisades and Eaton wildfires highlighted not only the significant financial losses but also the urgent necessity for better disaster readiness and response systems. The wildfires' aftermath showed the overwhelming extent of destruction, with more than 16,000 structures destroyed which surpassed  \$250 billion in economic loss \cite{guardian2025}. Beyond the immediate devastation, these wildfires revealed fundamental problems in evacuation planning, shelter accessibility, and resource allocation, underscoring the need for improvement to address the increasing difficulty of disaster management during periods of more frequent natural hazards. One of the most crucial concerns is that many evacuees faced multiple relocations due to the shortage of shelter capacity and insufficient accessibility to shelters \cite{nyt2025lafires}. The Palisades and Eaton Fires serve as a reminder that without significant improvements in emergency response systems--particularly in shelter accessibility--a disaster event will create prolonged displacement and hardship.

This study focuses specifically on the recent Palisades and Eaton wildfires, which exposed significant challenges in shelter accessibility during large-scale evacuations. Despite the severity of these events, spatial disparities in access to wildfire shelters have not been systematically examined. Accordingly, this research aims to evaluate the spatial accessibility of shelters during the Palisades and Eaton Fires, assessing whether the available facilities provided sufficient capacity to accommodate evacuees. It also investigates the equity of shelter access across different regions and analyzes how geographic variations in accessibility were influenced by traffic congestion during the evacuations. Furthermore, the study identifies potential new shelter locations to enhance overall accessibility and promote more equitable shelter distribution across all evacuation zones. Findings from this research will inform future shelter siting and emergency planning, supporting more effective protection and resource allocation for displaced populations during future wildfire disasters.

This research is intended to answer the research questions below.
\begin{itemize}

    \item[\textbf{Q1.}] To what extent was shelter accessibility effective during the peak of the Palisades and Eaton Fires—two of the most destructive wildfires that occurred in close proximity to densely populated residential areas—and did the available shelter capacity adequately accommodate the large number of displaced residents?
    
    \item[\textbf{Q2.}] What were the travel times to the nearest shelters for evacuees during the Palisades and Eaton Fires, and how did severe traffic congestion and road network disruptions influence spatial accessibility and exacerbate disparities in reaching safe shelters?
    
    \item[\textbf{Q3.}] How did shelter accessibility vary across the affected regions of the Palisades and Eaton Fires, and to what extent did spatial disparities emerge in terms of equitable access to emergency shelters among nearby residential communities?
    
    \item[\textbf{Q4.}] How could the strategic placement of additional shelters within the fire-affected areas improve overall accessibility and promote equitable, timely shelter access for residents living in high-risk and densely populated zones during wildfire emergencies such as the Palisades and Eaton Fires?

\end{itemize}

%% file: sections/related_work.tex
Related studies on wildfire evacuations and shelter accessibility fall into three main directions. The first explores the ongoing debate on whether evacuation always enhances safety or, in some cases, may inadvertently increase risk. The second focuses on the difficulties people face when trying to reach shelters located in regions with constrained space. The third extends its analysis beyond wildfires to measure shelter accessibility throughout multiple disaster situations.

\subsection{Evacuation vs. Shelter-in-Place (SIP) Strategies}
Surviving wildfires entails two principal strategies which include evacuation and shelter-in-place (SIP) \cite{cova2009protective}. Experts generally agree that prompt evacuation is the most effective strategy for reducing fatalities and injuries \cite{TCEP2025}. When adequate time is available, moving out of the fire affected area can give the greatest chance of survival \cite{cova2011modeling}. However, when active flames block escape routes or the fire advances too quickly, evacuation can potentially be riskier than sheltering in place due to the risk of encountering active flames \cite{Shurtz2025WildfireEvacuation}. In such cases, SIP serves as a last-resort option, potentially providing a higher chance of survival than taking the dangerous evacuation route. This strategy includes seeking refuge in well-prepared, fire-resistant structures \cite{handmer2005staying} or designated safety zones, such as cleared areas with minimal hazards \cite{baxter2004travel, SafeHomeWildfireGuide}. In extreme situations, people should seek refuge in water bodies to reduce heat exposure \cite{cova2011modeling}. Cova et al. \cite{cova2011modeling} emphasized that no single strategy guarantees safety in wildfire-prone regions, but evacuation remains the most effective option whenever it is feasible.

\subsection{Social and Geographic Inequities in Shelter Accessibility}
Research shows that social and geographic inequalities in access to emergency shelters limit disaster response effectiveness and place vulnerable populations at greater risk. Ermagun et al. noted that the elderly, individuals with disabilities, and Hispanic communities are at greater risk due to limited shelter access \cite{ermagun2024compound}. Moreover, geographic disparities create extra challenges for suburban and rural areas, which often struggle with sparse and insufficient emergency facilities \cite{ermagun2024compound, yang2023spatial, marolla2025california}.

Beyond existing shelter limitations, road network constraints create additional challenges to evacuation efforts. Evacuees’ ability to reach safety is hindered by sudden road blockages and fire-related damage to transportation networks \cite{dye2021evaluating, zehra2024systematic, fraser2022wildfire}. The presence of these barriers  not only delays evacuation times but also increases the likelihood of individuals becoming trapped within dangerous zones. As a result, effective evacuation planning and shelter accessibility are critical for mitigating wildfire risks and ensuring equitable access to emergency resources \cite{zehra2024systematic}.

\subsection{Spatial Accessibility and the Floating Catchment Area (FCA) Framework}
Beyond general evacuation concerns, scholars have recently noted that spatial accessibility has become an essential consideration for researchers working on emergency evacuation planning \cite{zehra2024systematic}. While accessibility methodologies have been extensively applied in healthcare research, including studies on access to hospitals, physicians, and preventive care services \cite{luo2009enhanced, park2023daily, kang2020rapidly, kang2022spatial, guagliardo2004spatial, luo2003measures, kang2023covid}, research on shelter accessibility during disaster evacuations has received less attention. In terms of shelter accessibility research, Floating Catchment Area (FCA) framework--including the Two-Step Floating Catchment Area (2SFCA) method, its various modifications, and enhanced versions--has been used to assess spatial accessibility in disaster scenarios, though not specifically for wildfires \cite{zhu2018improved, su2021using, zhang2023spatial, liang2023spatial, ding2022study, yang2023spatial, xu2025measuring}. These approaches combine essential elements, such as road network travel speed (traffic) with shelter capacity (supply) and detailed population (demand) data to analyze spatial access to emergency shelters effectively.

\subsection{Research Gaps in Wildfire Shelter Accessibility}
Although studies on emergency shelters' spatial accessibility have increased, there are still few that are especially focused on wildfires. For example, Ermagun and Janatababi \cite{ermagun2024compound} measured accessibility to wildfire shelters but did not account for the dynamic interactions among supply (shelters), demand (population), and travel time--key factors incorporated in the FCA framework. Furthermore, the majority of current studies evaluate spatial accessibility to urban emergency facilities such as shelters, fire stations, and emergency centers (e.g., \cite{zhu2018improved, zhao2017planning, unal2016gis, mandalapu2024evaluating, kiran2020measuring}), while often overlooking spatial accessibility in wildfire-prone regions. The populations in these areas are typically scattered throughout forested landscapes and have limited road infrastructure, which complicates emergency response efforts and shelter accessibility. Filling these research gaps will play a crucial role in enhancing wildfire disaster preparedness and ensuring equitable shelter access.

This research contributes to the field by analyzing spatial accessibility to shelters. Building upon prior research--where studies on shelter accessibility in disaster evacuations remain relatively scarce--this research applies advanced spatial accessibility methods to measure shelter accessibility during the Palisades and Eaton Fires. Specifically, the Enhanced Two-Step Floating Catchment Area (E2SFCA) method is employed to evaluate both supply-side factors (e.g., shelter capacity, available facilities) and demand-side dynamics (e.g., evacuee population distribution, mobility constraints). By integrating these critical factors, this study aims to offer new insights into improving shelter accessibility and emergency response strategies for future wildfire events.

%% file: sections/study_area.tex
\begin{figure}[tbp]
    \centering
    \includegraphics[width=1\linewidth]{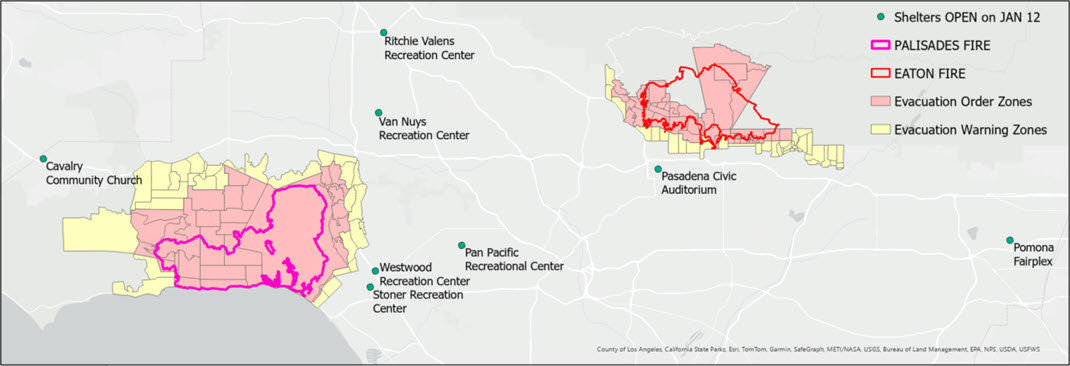}
    \caption{Study area}
    \label{fig:study_area}
\end{figure}

Our study area covers the regions where the Palisades and Eaton fires occurred (Figure \ref{fig:study_area}). Data on wildfire perimeters, evacuation orders, warning zones, and shelter statuses were collected via web crawling from CAL FIRE's website (fire.ca.gov), which provided frequent updates during active fire periods. According to their live updates, the Palisades Fire ignited around 10:30 a.m. in the Pacific Palisades neighborhood and expanded to over 11,000 acres by mid-January 8, eventually consuming approximately 23,713 acres. In contrast, the Eaton Fire ignited later on January 7 in the evening at Eaton Canyon near Altadena and expanded rapidly overnight to about 10,600 acres by January 8, eventually burning approximately 14,021 acres by January 16. CAL FIRE's live updates indicated that the Eaton Fire's perimeter peaked at 23,713.39 acres on January 12. However, later reports published on January 23, 2025, revised the burned area to 23,448 acres, with notes indicating it is under investigation \cite{CALFIRE:top20destructiveCAWildfires}. Both fires were fully contained on January 31, 2025.

Figure \ref{fig:study_area} illustrates the locations of each wildfire along with the corresponding evacuation and warning zones and shelter information as of January 12, 2025, the day we focus on examining shelter accessibility. We selected January 12 because, according to CAL FIRE's live reports, it was the date when the active wildfire expanded to its largest size. Additionally, this date represents an early stage of fire progression when displaced individuals were experiencing significant chaos due to quickly spreading fires that had begun only a few days earlier. At that time, many evacuees did not have clear information about the fire's trajectory and the extent of its impact, which made shelter accessibility an important concern.

%% file: sections/data.tex
\subsection{Shelter Capacity}

We gathered the locations, addresses, and overnight counts of available shelters as of January 12, 2025, from the Cal OES News website \cite{calOES2025}. The capacity of shelters open on January 12, 2025, as well as other potential shelters used in this study, was collected from the website, National Shelter System Facilities (NSSF) \cite{HIFLD_NSSFacilities}. The dataset includes 70,317 facilities across the U.S. that can accommodate individuals during an evacuation. While the downloaded shelter data included evacuation capacity information, this information was not consistently available for all shelters. Even though there were 11 shelters that remained open throughout the active Palisades and Eaton fires (as shown in Table \ref{tab:shelter_info}), capacity information was available for only six of them. For the remaining five shelters, we estimated capacity by assessing the available indoor space using various online sources. For instance, the Van Nuys/Sherman Oaks Recreation Center was found to have approximately 15,000 square feet of space, based on information obtained from a website \cite{paulmurdocharchitects}. We then calculated the usable shelter space according to the Federal Emergency Management Agency
 (FEMA) guidelines.

According to FEMA’s Guidelines for Mass Care Shelter Operations (FEMA P-785), planners must exclude non-occupiable areas—such as restrooms, kitchens, hallways, and administrative spaces—when calculating available shelter capacity \cite{FEMA_P785}. Following these recommendations, we assumed that approximately 70\% of each facility’s total floor area was suitable for occupancy. Furthermore, to ensure a comfortable and safe shelter environment, we allocated 100 square feet per person. This estimate exceeds the Centers for Disease Control and Prevention or CDC’s shelter assessment guidance, which recommends a minimum of 60 square feet per person in evacuation shelters \cite{CDC_ShelterAssessment}.
However, FEMA further advises that shelters should provide one toilet for every 20 people and one shower for every 25 people. Since no detailed information was available regarding the number of toilets or showers in each shelter facility, we adopted a more conservative estimate of 100 square feet per person to account for potential limitations. A detailed breakdown of how capacity was estimated for each of the five shelters is provided in Section~\ref{sec:shelters}. Table \ref{tab:shelter_info} shows all eleven shelter locations and capacities during the Palisades and Eaton wildfires. For facilities where capacity information is available on the website \cite{HIFLD_NSSFacilities}, the value is recorded in the capacity column. For facilities where capacity data is unavailable, the estimated capacity is provided in a separate column as estimated capacity.

\begin{table}[tbp]
    \centering
    \caption{Shelter locations and capacities during the Palisades and Eaton wildfires}
    \begin{tabular}{|c|c|c|}
        \hline
        \textbf{Location} & \textbf{Capacity} & \textbf{Estimated Capacity}\\
        \hline
        Ritchie Valens Recreation Center & 356 & \\
        \hline
        Van Nuys/Sherman Oaks Recreation Center & & 105\\
        \hline
        Calvary Community Church & & 797 \\
        \hline
        Pan Pacific Recreation Center & 598 & \\
        \hline
        Westwood Recreation Center & 855 & \\
        \hline
        Stoner Recreation Center & 350 & \\
        \hline
        Pasadena Civic Auditorium & & 910 \\
        \hline
        Pomona Fairplex & & 1056 \\
        \hline
        Glendale Civic Center & & 202 \\
        \hline
        Lanark Recreation Center & 100 & \\
        \hline
        El Camino Real Charter High School & 1100 & \\
        \hline
    \end{tabular}
    \label{tab:shelter_info}
\end{table}
The first eight shelters listed in Table \ref{tab:shelter_info} were open on January 12th. The three shelters at the bottom, along with additional facilities from  \cite{HIFLD_NSSFacilities}, were included to evaluate how accessibility could be improved by incorporating more potential shelters---one of the key objectives of this study, as outlined in the introduction.

\subsection{Population} 

\begin{figure}[tbp]
    \centering
    \includegraphics[width=1\linewidth]{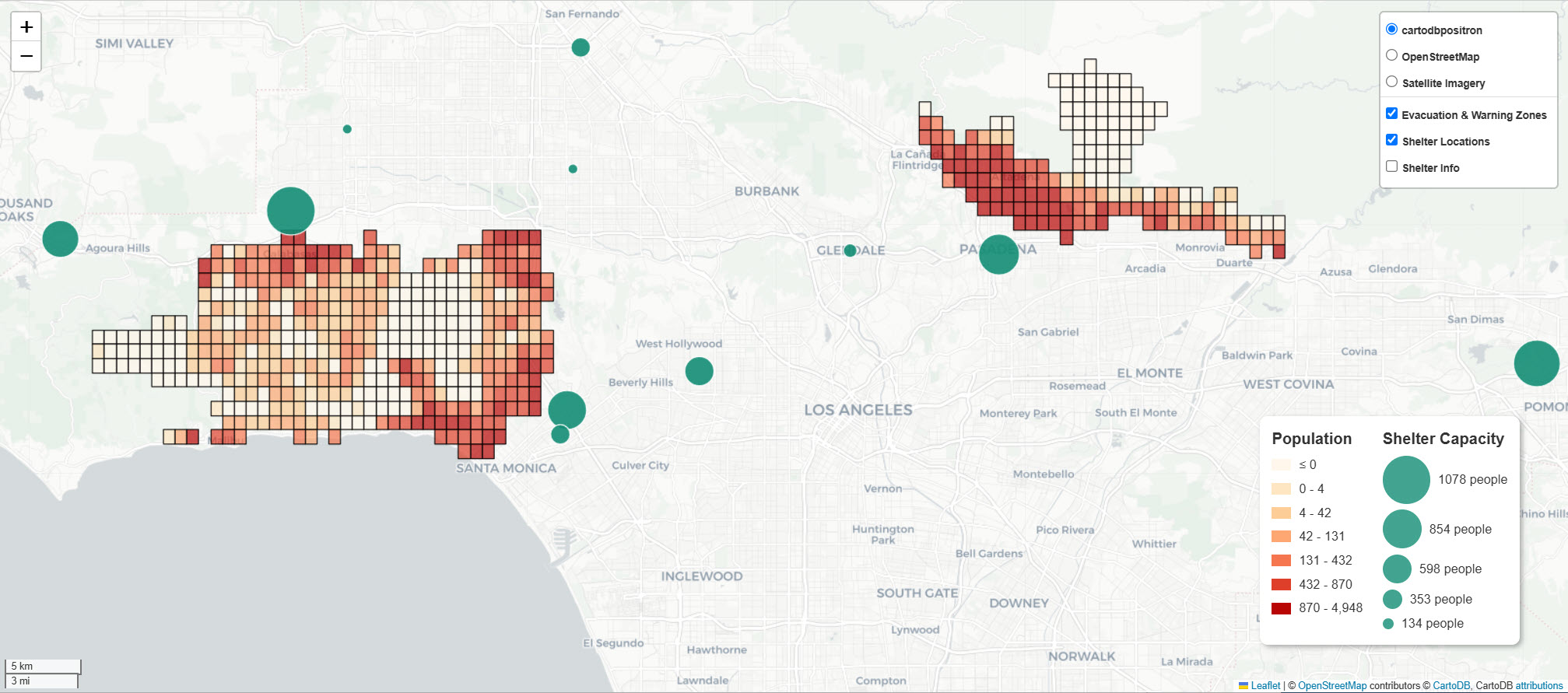}
    \caption{Population distribution based on LandScan Global 2023 data. \textit{Interactive version available at}: \url{https://sites.google.com/view/palisades-and-eton-fires/figure-2}}
    \label{fig:landscan}
\end{figure}

Initially, we considered using American Community Survey (ACS) population estimates at the census block group (CBG) level, the smallest geographic unit with available population data. However, CBG-based estimates assume uniform population distribution within each polygon, which is problematic in regions like Palisades and Eaton, where large portions are forested and sparsely populated. In some areas, CBGs cover vast mountainous regions, leading to inaccurate estimates---for instance, in a certain CBG, 90\% of the land is uninhabited, yet the data assumes even distribution across both populated and unpopulated areas. To address this issue, we used LandScan  \cite{LandScan}, which offers a higher resolution than CBGs and more accurately represents population distribution by avoiding assignments to forests and mountainous areas, as identified through satellite imagery. LandScan provides 1 km resolution grid data, representing a 24-hour average ambient population, estimated using a remote sensing-based global modeling and mapping approach \cite{LandScan}. LandScan also was used in previous accessibility studies \cite{cai2023transport, luqman2021geospatial}. For insights on how population distribution from different datasets, including LandScan, affects spatial accessibility evaluations, see \cite{tan2021effect}. Figure \ref{fig:landscan} shows the population distribution in Palisades and Eaton fire evacuation zones and the capacity of eleven shelters listed in Table \ref{tab:shelter_info} . Land cover (forest or residential) can be examined by switching base layers to OpenStreetMap or satellite imagery in the online map (link in the figure caption).

\subsection{Road Network}
The road data were collected  from OpenStreetMap (OSM) \cite{haklay2008openstreetmap}, which are a widely used dataset to compute travel cost in previous accessibility studies \cite{Liu2022generalized, luqman2021geospatial, kang2022spatial, kang2020rapidly, park2023daily}. We retrieved the OSM road network dataset using the Python package OSMnx \cite{boeing2017osmnx}, which allows for the downloading and analysis of street networks from OpenStreetMap. Given the sparse road networks, particularly in mountainous regions, we included all drivable roads, including service roads such as alleys and parking lot roads, to ensure comprehensive coverage. The road network downloaded in our study, as shown in Section~\ref{sec:road_network},
is extensive enough to cover wildfire-related evacuation routes surrounding shelters. The final road network dataset we downloaded contained 701,470 line features.

%% file: sections/method.tex
\subsection{Accessibility Measurement}
\label{sec:accessibility}

We applied the E2SFCA method to assess spatial accessibility between supply locations (i.e., shelters) and demand (i.e., population within each grid cell located in evacuation and warning zones). The E2SFCA method \cite{luo2009enhanced} builds upon the original 2SFCA method developed by \cite{luo2003measures}. It has been widely adopted in numerous studies, particularly for evaluating accessibility to healthcare services \cite{kang2022spatial, kang2020rapidly, park2023daily}. The E2SFCA method calculates accessibility in two steps, considering supply (shelter capacity), demand (population in each grid cell), and the travel cost between supply and demand locations, measured in time or distance. A key feature of this method is its incorporation of distance decay, which assumes that locations closer to a supply point have a greater influence, whereas those farther away have a lesser impact\cite{park2021review}. As the name suggests, the method consists of two steps. 

In the first step, for each shelter location $j$, search all population locations ($k$) in each grid that are within a threshold travel time ($t_0$) from location  $j$ (this is the catchment of shelter location $j$ or catchment $j$), and compute the weighted shelter-to-population ratio, $R_j$, within the catchment area as follows:

$$
R_j=\frac{S_j}{\sum_{k\in{t_{kj}\leq t_0}}{P_k W_{kj}}},
$$
where:
\begin{itemize}
    \item $R_j$ is the supply-to-demand ratio at shelter location j, representing the shelter capacity relative to the population it serves.
    \item $S_j$ is the degree of supply at shelter location $j$, measured as the shelter's capacity (e.g., number of evacuees it can accommodate).
    \item $P_k$ is the degree of demand at location $k$, measured as the population in grid cell $k$ whose centroid falls within the catchment $j$ (i.e., where $t_{kj} \leq t_0$).

    \item $t_{kj}$ is the travel time between population grid cell $k$ and shelter location $j$
    \item $t_0$ is the travel time threshold, set to 2 hours in this study. See the Section~\ref{sec:accessibility} for the rationale behind selecting the 2-hour threshold.

    \item $W_{kj}$ is a Gaussian-weighted distance decay function applied to the distance between the population grid cell $k$ and the shelter location $j$.
\end{itemize}

In the second step, each population grid $i$ searches for shelter locations $j$ within its catchment area and sums up $R_j$ values of these shelter  locations. The equation for Step 2 is defined as follows:
$$
A_i=\sum_{j\in \left\{ t_{ij} \leq t_0 \right\}}{R_j W_{ij}},
$$
where:
\begin{itemize}
    \item $A_i$  is the accessibility measure at location $i$.
    \item $R_j$  is the supply-to-demand (shelter-to-population) ratio at shelter location $j$ which falls within the catchment area centered on population location $i$ (i.e., where $t_{ij}\leq t_0$).
    \item $t_{ij}$ is the travel time between population grid cell $i$ and shelter location $j$
    \item $W_{ij}$ is the same Gaussian distance weights used in Step 1.
    \item $t_0$ is the \textit{travel time threshold}
\end{itemize}

The Spatial Access module in PySAL (The Python Spatial Analysis Library) \cite{pysalaccess} includes an implementation of the E2SFCA method, available through the function {\texttt{Access.enhanced\_two\_stage\_fca}} \cite{pysal2024enhanced2sfca}. One of the input parameters required for this function is the weight function $W_{ij}$, as defined in the above equations. Users can specify either a step function or a Gaussian weight function with a specified width ($\sigma$). In this study, we applied a Gaussian weight function with $\sigma$  = 30. The rationale for selecting this value is discussed in Section~\ref{sec:accessibility}.

\subsection{Computing Travel Cost}
\label{sec:travel_cost}
Travel time was computed based on the length and maximum speed of road segments within street networks extracted from OSM. As indicated in the equation above, the accessibility score is influenced not only by supply and demand but also by travel cost, which plays a crucial role in shaping access. The OSM street network includes road classification, segment length, and maximum speed, all of which vary based on road type. For instance, motorways, secondary roads, and service roads have maximum speeds of 65 mph, 35 mph, and 15 mph, respectively. When the maximum speed was unavailable for a road segment, we imputed a value based on the mean maximum speed value for edges categorized by highway type, using a function provided by the OSMnx Python package \cite{boeing2017osmnx}. Using the same package, we also computed travel times by leveraging the maximum speed and segment length for each road or line segment. Additionally, to compute the shortest paths from each demand location to all shelter locations, we employed Dijkstra's algorithm, a single-source shortest path algorithm, where the weight of an edge in the network is an estimated travel time from the starting vertex to the ending vertex of the edge. The travel cost, essential for measuring accessibility, was estimated by computing travel times along the shortest path from each demand location (i.e., each grid cell in Figure \ref{fig:landscan}) to each shelter location. In addition, when computing spatial accessibility under traffic congestion \cite{latimes2025palisades}, a maximum travel speed of 10 kph was assigned to the road network within a five-kilometer buffer zone around the Palisades and Eaton evacuation areas shown in Figure \ref{fig:landscan}. This assumes that vehicles could not travel at the designated road speeds and are instead limited to an average of 10 kph due to severe traffic jams and gridlock \cite{latimes2025palisades}.

\subsection{Measuring Disparities in Accessibility}
\label{sec:disparities}
We employ Gini coefficients to measure regional disparities in spatial accessibility to shelters. The Gini coefficient, a classical tool commonly used to measure income inequality, is also widely applied in evaluating the degree of imbalance in access to public service facilities \cite{harada2012study, nakamura2017potential, su2021using, wang2022assessing, sun2024spatial, guo2019accessibility, tahmasbi2019multimodal, shi2020urban}. The Gini coefficient is an effective measure for our study because it helps evaluate how evenly shelter access is distributed. Its values range from 0 to 1, where a lower Gini coefficient indicates more equitable access to shelters, while a value of 1 represents complete inequality. Specifically, a Gini coefficient of 0 reflects a perfectly uniform distribution, whereas a value of 1 indicates full concentration or extreme disparity.

The Gini coefficients is given by \cite{tahmasbi2019multimodal}: 
$$
G=1-\sum_{i=0}^{n}{(X_{i+1}-X_{i})(Y_{i+1}+Y_{i})},
$$
where $X_i$ denotes the cumulative proportion of the population within the population grid used for accessibility calculation, and $Y_i$ represents the cumulative proportion of accessibility to shelters.

\subsection{Workflow of This Study}
\label{sec:workflow}

\begin{figure}[tbp]
    \centering
    \includegraphics[width=1\linewidth]{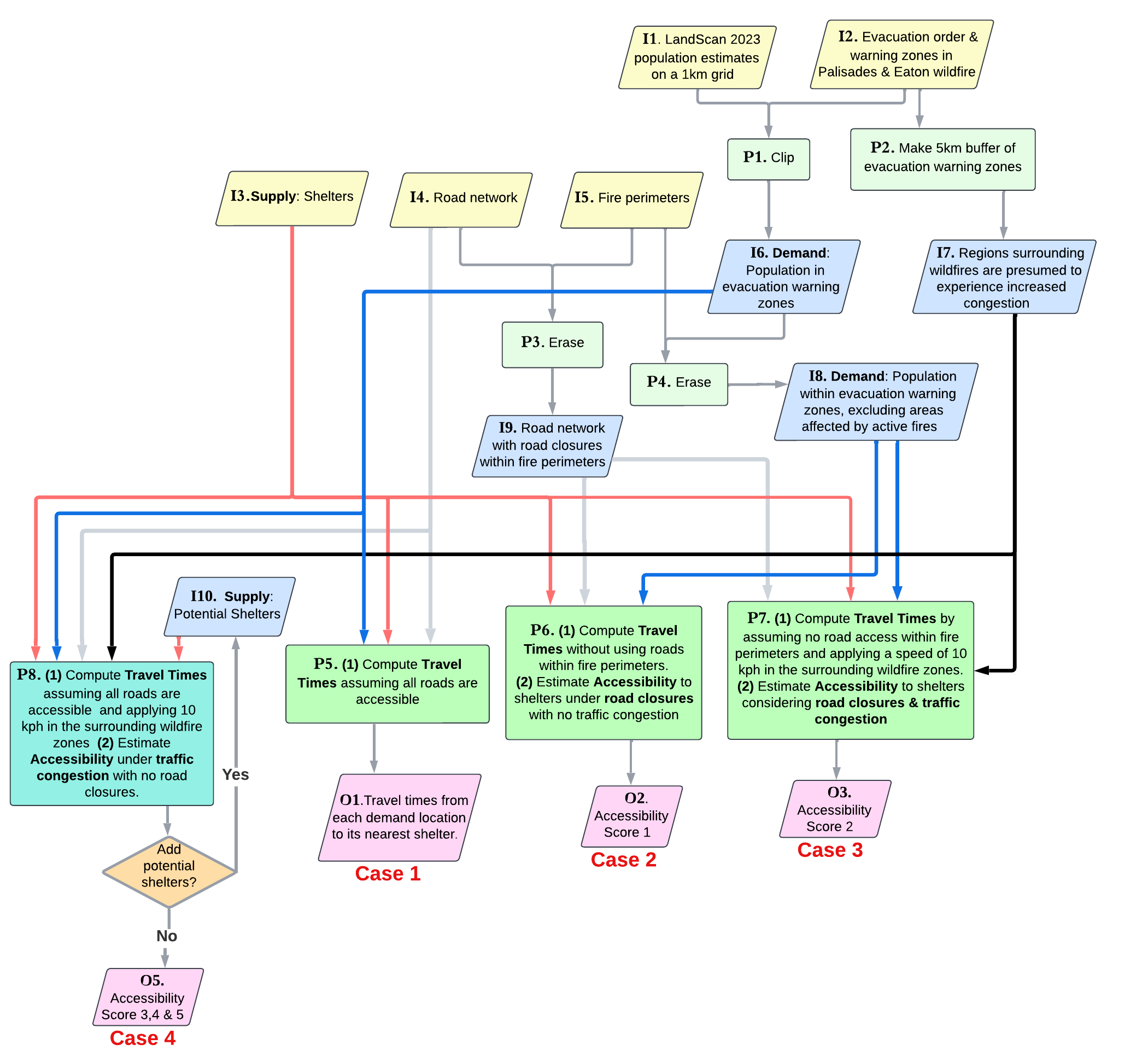}
    \caption{A diagram depicting work flow of this study}
    \label{fig:workflow}
\end{figure}

To illustrate the methodology employed in this study, Figure \ref{fig:workflow} shows the complete workflow used in this study to assess shelter accessibility during wildfire events. Inputs and outputs are represented by parallelograms--those labeled with an `I' indicate inputs, while those starting with an `O' represent outputs. Processes are shown as rectangles, each beginning with a `P' to denote a specific procedure or step.

The study workflow is structured into four distinct cases, each designed to evaluate accessibility under different spatial and situational conditions (Table \ref{tab:cases}). Across these cases, we vary three key components: shelter availability (supply)—including both operational and hypothetical shelters; demand definitions, based on evacuation orders, warning zones, and the inclusion or exclusion of areas affected by active fires; and travel costs, accounting for wildfire-induced traffic congestion and road closures.

Cases 1 through 3 represent real-world conditions during the peak of the wildfire crisis on January 12. As an illustrative example, Case 3 in Figure \ref{fig:workflow} corresponds to process P7, which models an event-driven scenario that departs from baseline accessibility assumptions by explicitly incorporating wildfire-related disruptions. Process P7 has two primary inputs: supply (I3), representing operational shelters, and demand (I8), defined as the population located within evacuation warning zones while excluding areas impacted by active fires. This exclusion is based on the assumption that residents do not remain in areas that are already burned or actively burning. Consequently, the demand region is constructed by erasing the fire perimeter zones (I5) from the evacuation warning zones (I6). The I6 layer is generated in P1 by clipping LandScan Global 2023 population estimates at a 1 km resolution (I1) with evacuation order and warning zones associated with the Palisades and Eaton wildfires (I2).

Process P7 further incorporates I7, which represents surrounding regions presumed to experience increased traffic congestion during the wildfire event. These regions are created in P2 by applying a 5 km buffer around evacuation warning zones derived from I2. Within these buffered areas, travel speed is assumed to be reduced to 10 kph, reflecting wildfire-related congestion. Together, P7 computes travel times by assuming no road access within fire perimeters and reduced travel speeds in surrounding wildfire-affected zones, enabling the estimation of shelter accessibility under conditions of road closures and severe traffic congestion. The rationale for selecting a 5 km buffer to delineate congestion zones and a travel speed of 10 kph is discussed in detail in Section~\ref{sec:workflow}.

In contrast, Case 4 examines hypothetical scenarios aimed at improving shelter accessibility under extreme evacuation demand. On January 12, a substantial mismatch existed between available shelter capacity and the population residing in evacuation zones. The eight operational shelters available on that day (the top eight listed in Table \ref{tab:shelter_info}) had a combined capacity of 5,224 people, whereas LandScan Global 2023 estimates indicate that 44,348 individuals were located in evacuation order zones and 42,263 in warning zones, totaling 86,611 people at risk. As a result, approximately 81,387 individuals lacked access to designated shelters. This pronounced disparity between shelter capacity and evacuation demand underscores the urgent need to expand shelter infrastructure during large-scale wildfire emergencies.

Motivated by this critical gap, Case 4 explores potential mitigation strategies by introducing hypothetical shelter sites. Candidate locations are drawn from the National Shelter System Facilities dataset \cite{HIFLD_NSSFacilities}, which identifies buildings that could potentially serve as shelters and provides information on their capacities.

\paragraph{Case 4.1: Capacity-based Approach}
--- The idea of this approach is that when there are too many shelters with small capacities, resources needed for evacuees (such as cots, blankets, food, and medical assistance) must be distributed across multiple locations. By prioritizing larger shelters, the approach aims to minimize logistical efforts and enhance the efficiency of resource allocation during emergencies. Our capacity-based greedy algorithm consists of two steps: filtering and refinement. In the filtering step, starting from the demand areas, the algorithm incrementally searches for the nearest shelters and adds them one by one until their combined capacity exceeds $k$ times the required demand. In our empirical analysis, when $k=2$, shelters within a reasonable distance are included. In the refinement step, the algorithm selects shelters in descending order of capacity from the candidates, starting with the largest, until the total capacity becomes sufficient to accommodate the estimated population in need.

\begin{algorithm}
\caption{Capacity-Based Shelter Selection}
\label{algo:capacity}
\KwIn{Set of demand areas $D$}
\KwIn{Set of shelters $S$ with capacities $C(s)$ for each $s \in S$}
\KwIn{Population demand $P_d$ for each $d \in D$}
\KwIn{Multiplier $k$}
\KwOut{Set of selected shelters $S_{\text{selected}}$}

$S_{\text{selected}} \gets \emptyset$\ \tcp*{Initialize the final list of selected shelters}

\ForEach{demand area $d \in D$}{
    $S_{\text{candidate}} \gets \emptyset$\ \tcp*{Temporary list of nearby shelters}
    $C_{\text{total}} \gets 0$\ \tcp*{Track cumulative capacity}

    \tcp{Filtering step: prioritize nearby shelters}
    Sort shelters $S$ by distance to $d$ in ascending order\; 

    \ForEach{ shelter $s \in S$ (in sorted order) }{
        Add $s$ to $S_{\text{candidate}}$\;
        $C_{\text{total}} \gets C_{\text{total}} + C(s)$\;
        \If{ $C_{\text{total}} \geq k \cdot P_d$ }{
            \textbf{break}\ \tcp*{Stop once buffer capacity (k × demand) is met}
        }
    }

    \tcp{Refinement step: prioritize larger shelters}
    Sort $S_{\text{candidate}}$ by capacity $C(s)$ in descending order\; 
    $C_{\text{refined}} \gets 0$\ \tcp*{Track capacity in refined selection}

    \ForEach{ shelter $s \in S_{\text{candidate}}$ }{
        Add $s$ to $S_{\text{selected}}$\;
        $C_{\text{refined}} \gets C_{\text{refined}} + C(s)$\;
        \If{ $C_{\text{refined}} \geq P_d$ }{
            \textbf{break}\ \tcp*{Stop once actual demand is met}
        }
    }
}

\Return $S_{\text{selected}}$\ \tcp*{Return selected shelters across demand areas}

\end{algorithm}

Algorithm~\ref{algo:capacity} shows the detail of the capacity-based approach as pseudo code. 
The algorithm begins by initializing an empty set to store the final selected shelters (Line~1). For each demand area (Line~2), a candidate list of nearby shelters is initialized (Line~3), and a total capacity counter is set to zero (Line~4). 
In the filtering step (Lines~5--10), the algorithm sorts all shelters by their distance to the current demand area (Line~5), and iteratively adds them to the candidate list until the combined capacity exceeds $k \cdot P_d$ (Lines~6--10). This buffer-based threshold ensures that a sufficient number of shelters are considered without exceeding a reasonable distance.
Once the candidate set is built, the refinement step begins by sorting it in descending order of shelter capacity (Line~11). Then, starting from the largest shelters, the algorithm selects shelters into the final output set until the actual population demand is met (Lines~12--16).
Finally, the set of all selected shelters across all demand areas is returned (Line~17).

\paragraph{Case 4.2: Distance-based Approach} --- The intuition behind this approach is to provide evacuees with the nearest possible shelter locations, thereby reducing the time and distance needed to reach shelters. Our distance-based greedy algorithm is described as follows. Starting from the demand areas, the algorithm performs a buffer-based search, incrementally increasing the buffer distance (for example, from 1 mile to 2 miles, and so on). The algorithm continues expanding buffer distances until the total shelter capacity reaches the required demand. 

Algorithm~\ref{algo:distance} presents the distance-based approach in pseudo code. 
%This method aims to assign evacuees to the nearest feasible shelters by gradually increasing the search radius until demand is satisfied. 
The algorithm initializes an empty set to hold the final selection of shelters (Line~1). For each demand area (Line~2), it begins with an initial buffer radius $\delta$ (Line~5) and an empty candidate set (Line~3),  as well as a capacity counter (Line~4).
Within the while-loop (Lines~6--11), the algorithm identifies all shelters within the current radius (Line~7). Newly discovered shelters are added to the candidate list, and their capacities are accumulated (Lines~8--9). If the current total capacity is still below the required demand, the buffer is increased (Line~10), and the process repeats. Once the cumulative capacity meets or exceeds the demand, all candidate shelters found within the final buffer are added to the overall selection (Line~12). The algorithm proceeds to the next demand area and finally returns the complete list of selected shelters (Line~13).

\begin{algorithm}
\caption{Distance-Based Shelter Selection}
\label{algo:distance}
\KwIn{Set of demand areas $D$}
\KwIn{Set of shelters $S$ with capacities $C(s)$ and locations}
\KwIn{Population demand $P_d$ for each $d \in D$}
\KwIn{Initial buffer distance $\delta$}
\KwIn{Buffer increment $\Delta$}
\KwOut{Set of selected shelters $S_{\text{selected}}$}

$S_{\text{selected}} \gets \emptyset$\ \tcp*{Initialize the final list of selected shelters}

\ForEach{demand area $d \in D$}{
    $S_{\text{candidate}} \gets \emptyset$\ \tcp*{Temporary set of nearby shelters}
    $C_{\text{total}} \gets 0$\ \tcp*{Total capacity found so far}
    $r \gets \delta$\ \tcp*{Start with initial buffer radius}

    \While{$C_{\text{total}} < P_d$}{
        Find shelters $S_r \subseteq S$ within radius $r$ of demand area $d$\;
        \ForEach{$s \in S_r \setminus S_{\text{candidate}}$}{
            Add $s$ to $S_{\text{candidate}}$\;
            $C_{\text{total}} \gets C_{\text{total}} + C(s)$\;
        }
        \If{$C_{\text{total}} < P_d$}{
            $r \gets r + \Delta$\ \tcp*{Expand buffer and continue search}
        }
    }

    Add all shelters in $S_{\text{candidate}}$ to $S_{\text{selected}}$\;
}

\Return $S_{\text{selected}}$\ \tcp*{Return shelters selected for all demand areas}

\end{algorithm}

\newgeometry{left=1cm,top=1cm,left=2cm,bottom=2cm}
\begin{landscape}
\begin{table}[h]
    \centering
    \caption{Summary of the four case scenarios and procedures used in this study. The labels (e.g., I3, P5) correspond to the input, output and process labels shown in the diagram in Figure \ref{fig:workflow}.}
    \vspace{5pt}
    \small
    \begin{singlespace}
    \begin{tabular}{|l|>{\raggedright\arraybackslash}p{2.2cm}|>{\raggedright\arraybackslash}p{3.5cm}|>{\raggedright\arraybackslash}p{3cm}|>{\raggedright\arraybackslash}p{3cm}|>{\raggedright\arraybackslash}p{3cm}|>{\raggedright\arraybackslash}p{4cm}|}
    \hline
    \textbf{Case} & \textbf{Input: Supply} & \textbf{Input: Demand} & \textbf{Input: Road Network} & \textbf{Traffic Congestion} & \textbf{Process} & \textbf{Output} \\
    \hline
    Case 1 & I3. 8 shelters opened on Jan 12th & I6. All population in evacuation order \& warning zones & I4. Full road network & Not considered & P5. Compute travel time from each demand location to its nearest shelter & Figure \ref{fig:travel_time}: Travel times from each grid cell to its nearest shelter \\
    \hline
    Case 2 & I3. 8 shelters opened on Jan 12th & I8. Population within evacuation order \& warning zones, excluding areas affected by active fires & I9. Road network excluding roads within fire perimeters & Not considered & P6. Compute accessibility using specified supply, demand, and road network & Figures \ref{fig:closure_access_score} \& \ref{fig:closure_no_congestion}: Accessibility with road closures \\
    \hline
    Case 3 & I3. 8 shelters opened on Jan 12th & I8. Population within evacuation order \& warning zones, excluding areas affected by active fires & I9. Road network excluding roads within fire perimeters & I7. Considered & P7. Compute accessibility using specified supply, demand, and road network with congestion & Figure \ref{fig:closure_congestion}: Accessibility with road closures and traffic congestion \\
    \hline
    Case 4 & Varied hypothetical shelters & I6. All population in evacuation order \& warning zones & I4. Full road network & I7. Considered & P8. Compute accessibility using specified supply, demand, and road network with congestion & Figures \ref{fig:greedy_congestion} \& \ref{fig:distance_threshold_congestion}: Accessibility based on hypothetical shelters sized for all residents 
    
    Figure \ref{fig:open_congestion}: Accessibility to 8 shelters opened on Jan 12th \\
    \hline
    \end{tabular}
    \end{singlespace}
    \label{tab:cases}
\end{table}
\end{landscape}
\restoregeometry

The demand for shelter accessibility included the population in both evacuation order and warning zones to realistically represent evacuation behaviors observed in recent wildfire events. During wildfire events, the observed evacuation behavior showed that residents from both evacuation order and warning zones chose to evacuate even without official orders, which contradicts the widely held belief that only those in evacuation order zones would evacuate. During the Palisades Fire in January 2025, evacuation started before official orders were given, as people noticed that homes were already burning and traffic congestion had formed by the time they received official evacuation orders \cite{apnews2025palisades}. For this reason, many people from evacuation warning zones decided to leave their homes voluntarily during high-risk and fast-changing conditions, similar to residents in evacuation order zones. The unpredictable fire behavior and rapid spread forced residents of both zones to seek shelter before receiving official orders. Therefore, to assess shelter accessibility, we considered the population in both evacuation warning and evacuation order zones as the demand in the spatial accessibility analysis.

%% file: sections/results.tex
\subsection{Distance to nearest shelters}
\label{sec:dist_shelters}

\begin{figure}[tbp]
    \centering
    \includegraphics[width=1\linewidth]{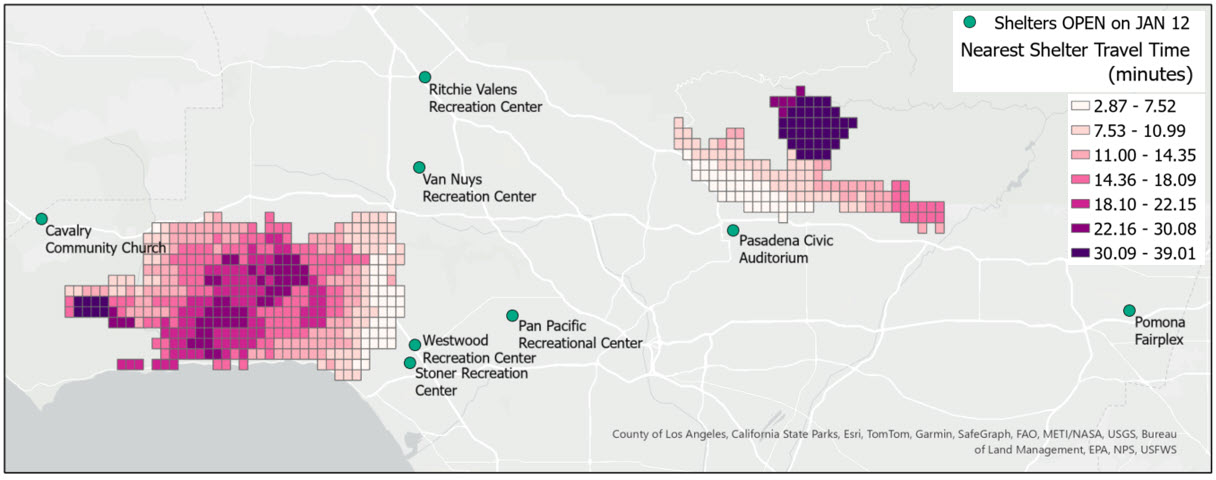}
    \caption{Travel times from each demand location (i.e., each grid cell) to its nearest shelter, based on Case 4 described in Table \ref{tab:cases} and illustrated in Figure \ref{fig:workflow}.}
    \label{fig:travel_time}
\end{figure}

General insights into travel times to shelters can be drawn from Figure \ref{fig:travel_time}. In this map, the color shading of each grid cell represents estimated travel times to the nearest shelters under conditions without road congestion. For the Palisades Fire, the southeastern regions within the evacuation order and warning zones generally have shelters located within approximately 3 to 8 minutes, indicating relatively easy access compared to other areas. In contrast, travel times from the inner mountainous regions can extend up to 40 minutes. For the Eaton Fire, the Pasadena Civic Auditorium is the closest shelter for residents in the evacuation order zones. Travel times to this location are generally up to 20 minutes, except for the mountainous areas. In these sparsely populated mountain regions, travel times are more than double those in residential zones. Figure \ref{fig:travel_time} illustrates travel times only to the nearest shelters whose capacities were significantly smaller than the demand as discussed in Section \ref{sec:workflow}; as a result, many evacuees had to travel to shelters farther away than the closest ones.

\subsection{Shelter accessibility under congested and uncongested traffic conditions}
\label{sec:access_traffic_congestion}

\begin{figure}[tbp]
    \centering
    \includegraphics[width=1\linewidth]{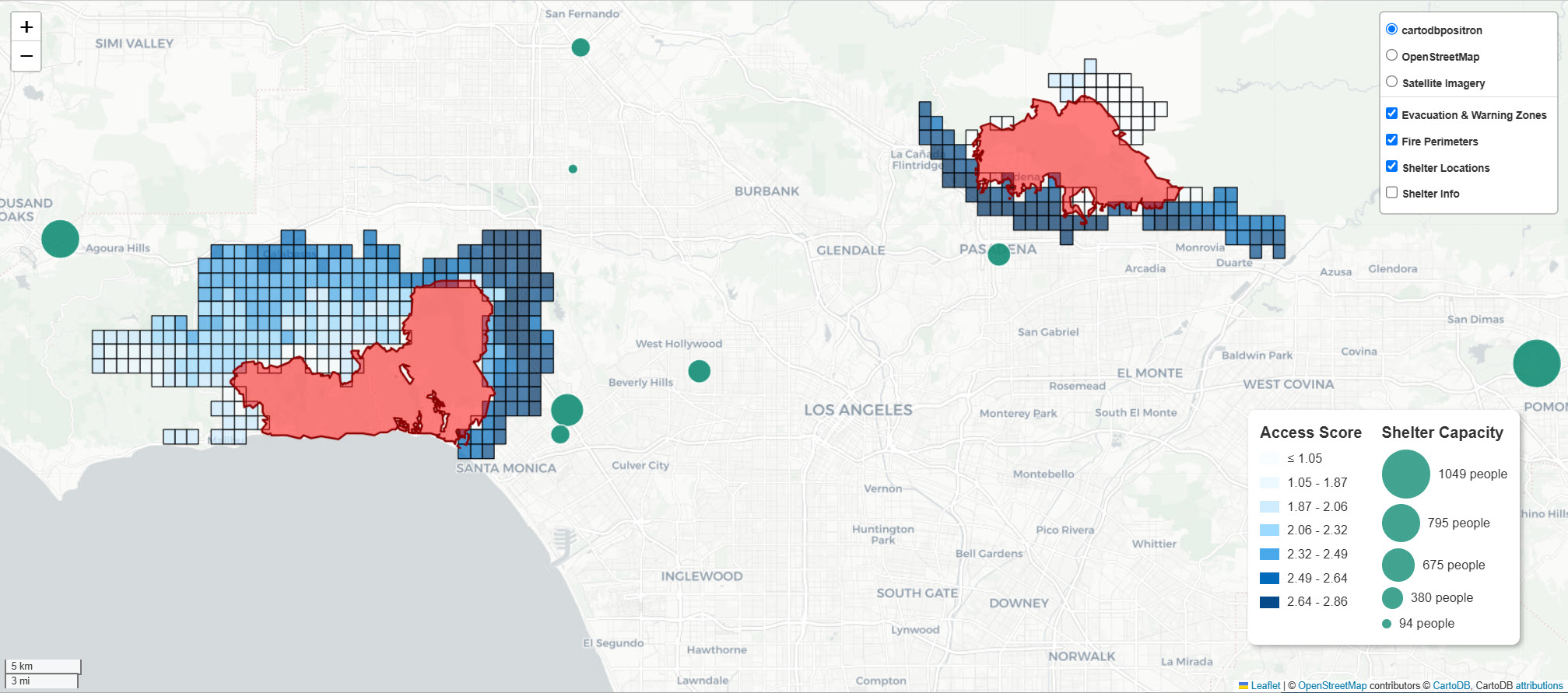}
    \caption{Shelter accessibility from evacuation order and warning zones on January 12, assuming road closures but no traffic congestion. Interactive version available at: \url{https://sites.google.com/view/palisades-and-eton-fires/figure-5}}
    \label{fig:closure_access_score}
\end{figure}

Figure \ref{fig:closure_access_score} presents spatial accessibility scores on January 12--the day the wildfire reached its largest extent, and eight shelters were operational. Roads within the active fire perimeters, marked in red, were considered inaccessible. It was assumed that all populations within these zones had already evacuated, so they were excluded from the demand calculations. Shelter capacities were also adjusted to reflect evacuees already accommodated. Accessibility scores on this day ranged from 0 to 2.86. In the Palisades evacuation zones, accessibility was generally higher on the eastern side toward Los Angeles, while lower scores were observed in mountainous areas and the southwestern portion of the zones. In the Eaton evacuation zones, scores ranged from 2.5 to 2.86 in southern residential areas, indicating relatively good access. However, some sparsely populated grids--particularly in the northern mountainous region--showed scores of zero due to road closures that blocked evacuation routes.

\begin{figure}[tbp]
    \centering
    \includegraphics[width=1\linewidth]{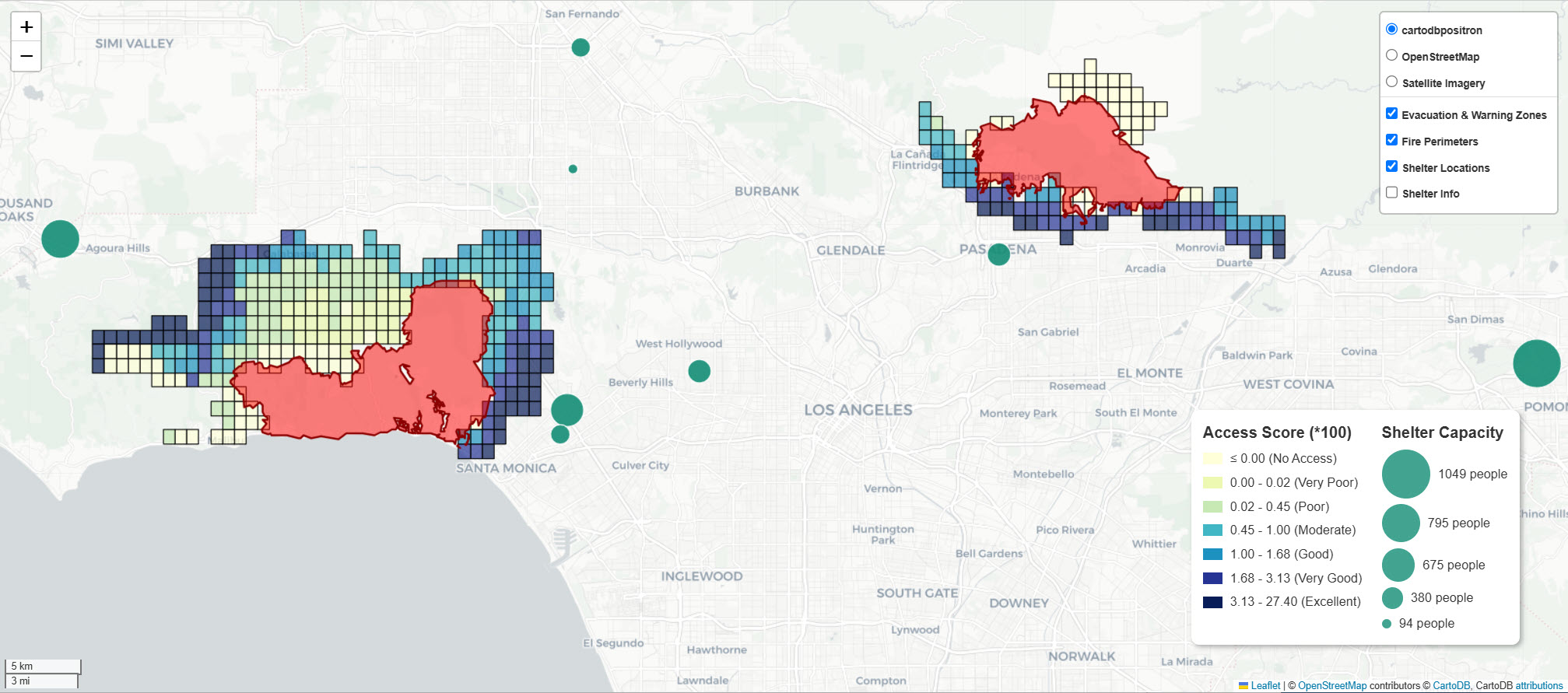}
    \caption{Shelter accessibility from evacuation order and warning zones on January 12, accounting for road closures and traffic congestion. Interactive version available at: \url{https://sites.google.com/view/palisades-and-eton-fires/figure-6}}
    \label{fig:closure_congestion}
\end{figure}

\begin{figure}[tbp]
    \centering
    \includegraphics[width=1\linewidth]{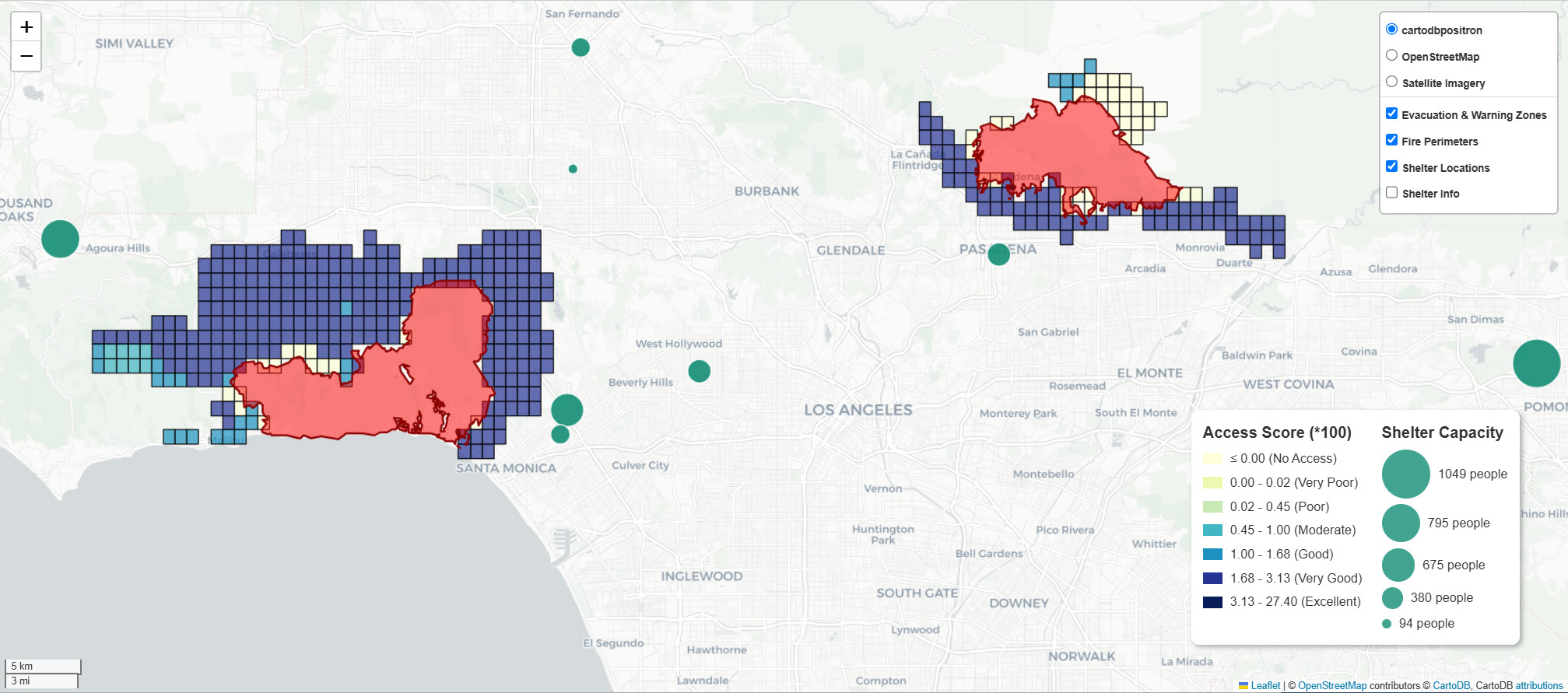}
    \caption{Shelter accessibility from evacuation order and warning zones on January 12, accounting for road closures but without traffic congestion. Interactive version available at: \url{https://sites.google.com/view/palisades-and-eton-fires/figure-7}}
    \label{fig:closure_no_congestion}
\end{figure}

Figures \ref{fig:closure_congestion} and \ref{fig:closure_no_congestion} build upon this baseline to examine the impact of traffic congestion on spatial accessibility. Figure \ref{fig:closure_congestion} reflects congested traffic conditions, as defined in Case 2 of Table \ref{tab:cases} and Figure \ref{fig:workflow}, where travel times were significantly increased due to congestion. In this scenario, not only were roads within the fire perimeters closed, but travel within a five-kilometer buffer zone around the evacuation areas was assumed to be extremely limited, with speeds reduced to just 10 kph.

Figure \ref{fig:closure_no_congestion} reclassifies the results from Figure \ref{fig:closure_access_score} using the same interval scheme as Figure \ref{fig:closure_congestion}, allowing for direct comparison between conditions with and without congestion. Both maps use harmonized class intervals--\textit{No Access}, \textit{Very Poor}, \textit{Poor}, \textit{Moderate}, \textit{Good}, \textit{Very Good}, and \textit{Excellent}--to depict differences in accessibility levels across scenarios.

The comparison between Figures \ref{fig:closure_congestion} and \ref{fig:closure_no_congestion} highlights the negative impact of traffic congestion on equitable access to shelters. Figure \ref{fig:closure_no_congestion}, which depicts accessibility without congestion, shows generally high access across most of the study area--except near active fire perimeters and in regions with sparse road networks, particularly in the southwestern Palisades zones and the mountainous areas of the Eaton evacuation zones. In contrast, Figure \ref{fig:closure_congestion}, which incorporates the effects of traffic congestion, reveals a substantial decline in accessibility. Only a few areas--such as narrow areas the northwest and southeast Palisades fire evacuated zones and a small region near the Pasadena Civic Auditorium near Eaton fire evacuated zones—maintain high accessibility scores. Most other areas fall into the \textit{Very Poor} to \textit{Moderate} accessible categories, indicating a widespread decline in shelter access under congested conditions.

This disparity is further supported by Gini coefficients, which quantify accessibility inequality. The coefficients for Figures \ref{fig:closure_congestion} and \ref{fig:closure_no_congestion} are 0.45 and 0.09, respectively, demonstrating a significant increase in inequality under congested traffic conditions.

\subsection{Shelter accessibility with hypothetical shelter locations}
\label{sec:access_shelter_locations}

As discussed in Section \ref{sec:workflow}, approximately 81,387 individuals were left without access to designated shelters. In this section, we explore strategies for the placement of additional shelters to accommodate the entire at-risk population and improve overall accessibility during wildfire emergencies. Specifically, we apply two approaches introduced in Section \ref{sec:workflow}--the Capacity-based Approach (Case 4.1) and the Distance-based Approach (Case 4.2). Both cases assume traffic congestion, with travel speeds reduced to 10 kph within a five-kilometer buffer zone surrounding the evacuation areas, consistent with the previous scenario.

\begin{figure}[tbp]
    \centering
    \includegraphics[width=1\linewidth]{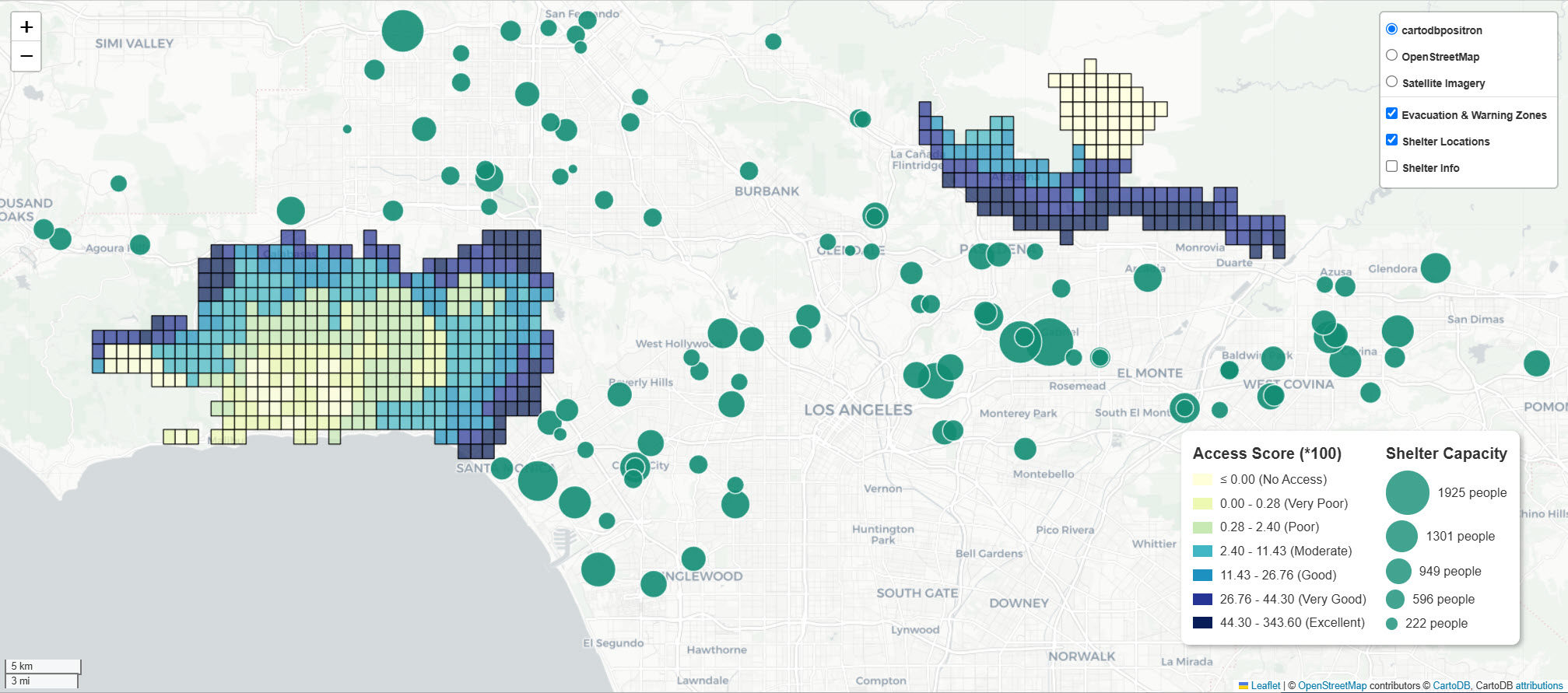}
    \caption{Shelter accessibility based on hypothetical shelters sized to accommodate all residents. A greedy algorithm prioritized shelters with the largest capacities until population needs were met. Interactive version: \url{https://sites.google.com/view/palisades-and-eton-fires/figure-8}}
    \label{fig:greedy_congestion}
\end{figure}

Figure \ref{fig:greedy_congestion} presents the results of the Capacity-based Approach, as outlined in Case 4.1 of Section \ref{sec:workflow}. In this scenario, the first eleven shelters used during the Palisades and Eaton wildfires were included, providing a combined capacity for 6,424 evacuees--significantly below the total at-risk population of 86,661 within the evacuation zones. To address this shortfall, the algorithm began from demand areas  (i.e., the Palisades and Eaton zones) and incrementally searched for additional potential shelters drawn from the National Shelter System Facilities dataset (\cite{HIFLD_NSSFacilities}) by expanding buffer distances outward. Shelters were added one by one until the combined capacity reached twice the estimated demand (173,222 people) to ensure sufficient coverage. This threshold was reached at approximately 13 kilometers from the evacuation zones. At that point, shelter selection was finalized by choosing facilities within the 13-kilometer buffer in descending order of capacity--starting with the largest--until the cumulative shelter capacity met or exceeded the population in need (86,661 people). Through this process, the selected shelters shown in Figure \ref{fig:greedy_congestion} were identified, offering a total combined capacity of 86,957 evacuees.

\begin{figure}[tbp]
    \centering
    \includegraphics[width=1\linewidth]{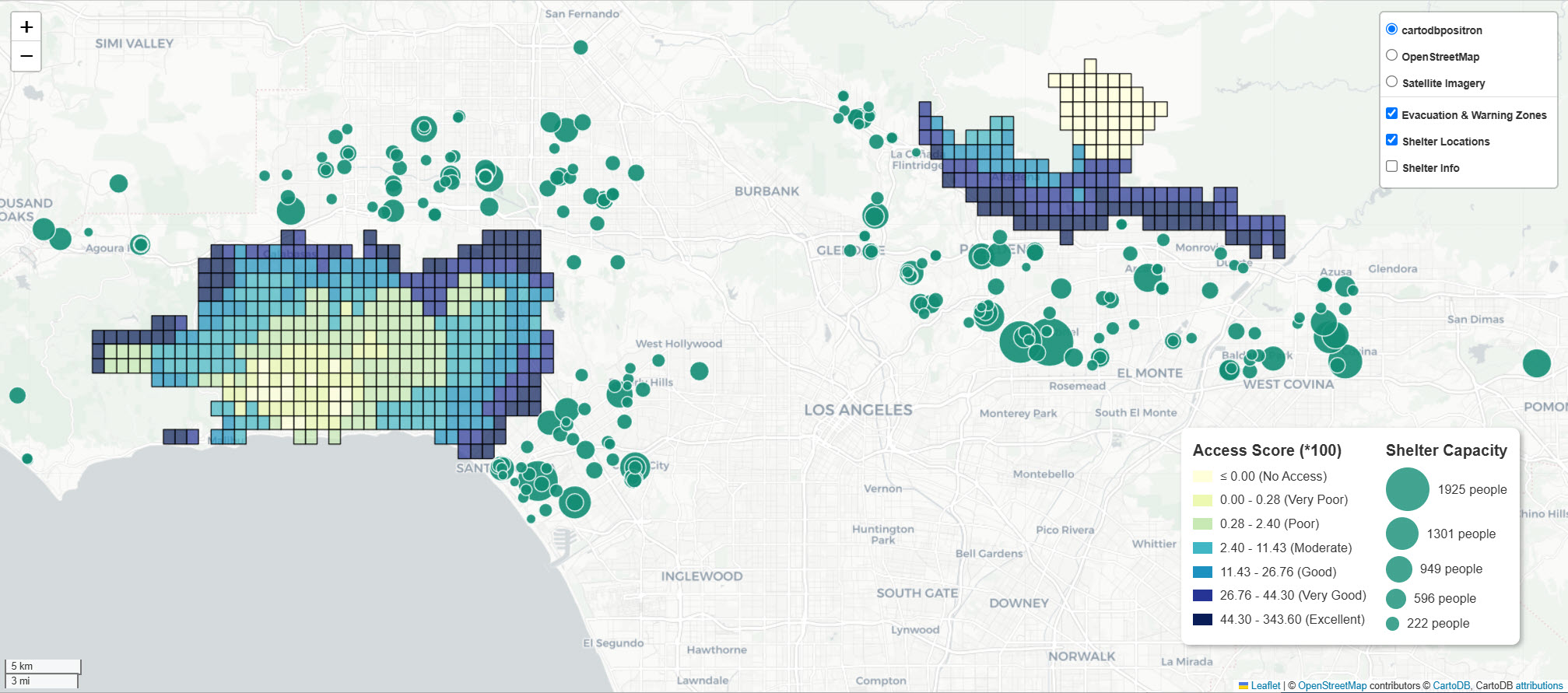}
    \caption{Shelter accessibility based on hypothetical shelters sized to accommodate all residents. Shelters were incrementally added in order of proximity until the entire population in evacuation and warning zones was covered. Interactive version: \url{https://sites.google.com/view/palisades-and-eton-fires/figure-9}}
    \label{fig:distance_threshold_congestion}
\end{figure}

Figure \ref{fig:distance_threshold_congestion} presents the results of the Distance-based Approach (Case 4.2). Similar to the Capacity-based Approach, this method began by including the eleven existing shelters, which together provided a total capacity of 6,424 evacuees. However, unlike the Capacity-based Approach, to accommodate the remaining population--40,092 individuals in the Palisades evacuation zones and 40,095 in the Eaton zones--a buffer-based search was conducted outward from each demand area. The algorithm incrementally expanded the buffer distance, adding the nearest available shelters until the cumulative shelter capacity, based on the National Shelter System Facilities dataset \cite{HIFLD_NSSFacilities}, met or slightly exceeded the required demand for each zone. The selection process concluded once the total shelter capacity reached 40,157 for the Palisades zone and 41,374 for the Eaton zone.

The resulting distribution of shelters is shown in Figure \ref{fig:distance_threshold_congestion}. Unlike the shelter pattern in Figure \ref{fig:greedy_congestion}, which focuses on maximizing total capacity regardless of location, the Distance-based Approach in Figure \ref{fig:distance_threshold_congestion} yields a more spatially clustered distribution, with shelters located closer to the evacuation zones. This outcome reflects the algorithm's emphasis on minimizing travel distance for evacuees by prioritizing proximity over shelter capacity.

\begin{figure}[tbp]
    \centering
    \includegraphics[width=1\linewidth]{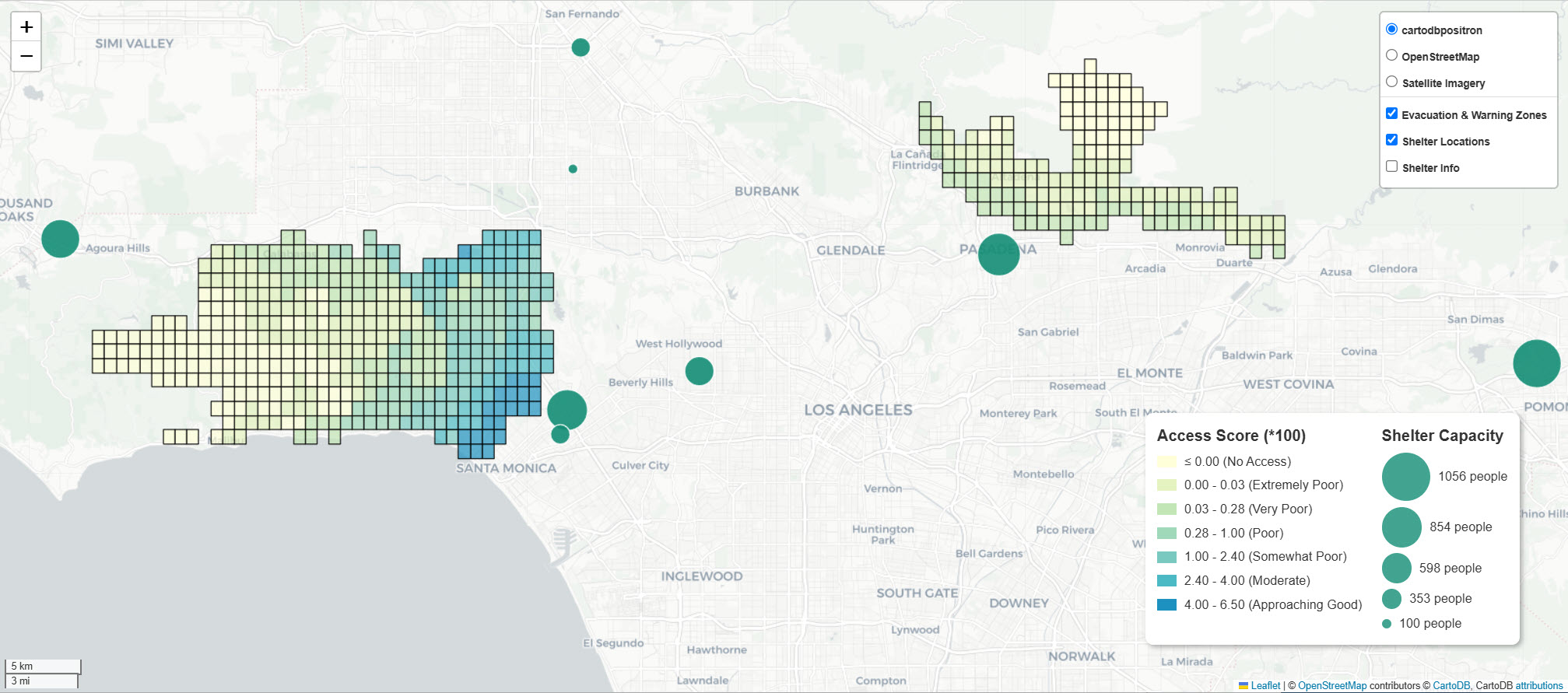}
    \caption{Shelter accessibility from evacuation order and warning zones on January 12, assuming full road network availability and traffic congestion. Interactive version: \url{https://sites.google.com/view/palisades-and-eton-fires/figure-10}}
    \label{fig:open_congestion}
\end{figure}

Shelter placements using the Capacity-based Approach (Figure \ref{fig:greedy_congestion}) and the Distance-based Approach (Figure \ref{fig:distance_threshold_congestion}) show greater improvements in accessibility compared to the actual eleven shelters opened on January 12, as shown in Figure \ref{fig:open_congestion}. Figure \ref{fig:open_congestion} presents the accessibility scores under the same traffic congestion scenario depicted in Figures \ref{fig:greedy_congestion} and \ref{fig:distance_threshold_congestion}. To ensure a fair comparison, the color scheme and class intervals in Figure \ref{fig:open_congestion} were aligned with those in Figures \ref{fig:greedy_congestion} and \ref{fig:distance_threshold_congestion}. As expected, the overall accessibility scores in the hypothetical shelter scenarios (Figures \ref{fig:greedy_congestion} and \ref{fig:distance_threshold_congestion}) are generally higher across the study area due to the increased number of shelters, compared to the case of eleven shelters in Figure \ref{fig:open_congestion}. When examining equity in shelter access across the three scenarios, the Gini coefficient is 0.34 for the Capacity-based Approach and 0.31 for the Distance-based Approach (Figure \ref{fig:distance_threshold_congestion}), while the original 11-shelter scenario (Figure \ref{fig:open_congestion}) shows a much higher Gini coefficient of 0.69. This indicates significantly greater disparities in shelter access under the actual 11-shelter scenario.

Between the two hypothetical strategies in Figure \ref{fig:greedy_congestion} and \ref{fig:distance_threshold_congestion}, the Distance-based Approach shows slightly better equity, as indicated by its lower Gini coefficient. This suggests that the Distance-based approach provides more equal access to shelters than the Capacity-based Approach. The likely reason is that it prioritizes the minimum travel distance and time for evacuees, while the Capacity-based Algorithm focuses on reducing the number of shelters required, emphasizing efficiency and logistical resource distribution.

%% file: sections/conclusion.tex
This research investigates four essential research questions focused on how accessible shelters were during the 2025 wildfires in Palisades and Eaton. First, the analysis showed that during the wildfire peak period shelter accessibility was inadequate because eight shelters with a total capacity of 5,224 served more than 86,000 evacuees, leaving the vast majority without access to formal shelter access. Second, evacuees needed to travel significantly varying length of time to reach the nearest shelters which often took more than 30–40 minutes from mountainous and remote areas during normal conditions but became even longer due to road closures and traffic congestion. Third, there were notable geographic disparities in shelter accessibility, particularly in isolated areas lacking sufficient road infrastructure. Lastly, the implementation of strategic shelter placement using the Capacity-based and Distance-based Approaches demonstrated substantial improvements in both shelter accessibility and equity in access by expanding shelter capacity and efficient allocation of shelter locations which provided actionable future emergency solutions.

The results from our study provide multiple policy implications for improving disaster preparedness and resilience in areas that frequently experience wildfires.

First, shelter placement is crucial in improving accessibility during wildfire evacuations. Our analysis demonstrates that accessibility can be significantly improved through strategic shelter placement. Utilizing the Capacity-based and Distance-based approaches, we proposed shelter placements to enhance access and bolster urban resilience in emergency situations. Our results suggest that data-driven planning approaches can enhance readiness and resilience against disasters in wildfire-prone regions. To achieve this, emergency management agencies need to develop region-specific plans that align shelter availability with projected displacement figures, using detailed population data. Additionally, state and county authorities should establish minimum shelter-to-population standards for high-risk regions to maintain sufficient emergency service capacity and equitable distribution of emergency resources.

Second, enhancing urban resilience to wildfires necessitates forward-thinking planning and strategic investment in shelter facilities. Proactive site designation and infrastructure investment are essential for achieving emergency readiness. Maintaining an up-to-date inventory of potential emergency shelter locations--such as schools and community centers--is essential for rapid deployment during crises. Vulnerable communities and geographically isolated areas need immediate infrastructure enhancements that promote equitable access to emergency shelters. These upgrades should prioritize compliance with the American with Disabilities Act (ADA), along with improvements to sanitation and ventilation systems. Equally important is the inclusion of spatial equity in emergency planning. Shelter siting should consider geographic distribution, road network limitations, and vulnerable population so that all communities--not just those in urban centers--have fair access to life-saving services. It should also be noted that collaboration between transportation agencies and emergency managers is vital to enhance urban resilience planning for disaster events. Integrating real-time traffic updates and evacuation simulations into preparedness plans will help reduce congestion and prevent access bottlenecks during evacuations.

Finally, key elements to building resilient communities include scenario-based preparedness drills together with forward-looking policies. Local authorities should routinely conduct exercises to test shelter accessibility under various hazard scenarios, using simulations like those developed in this study. Through routine preparedness drills, local planners and decision-makers can move from reactive approaches—where they respond only after a disaster happens—to proactive systems that are based on data-driven systems designed to ensure fair access to shelters for everyone. Given the increasing frequency and intensity of climate-driven disasters coupled with growing unpredictability, it is essential to view shelter accessibility not only as a logistical necessity but also as a matter of environmental justice. This research offers a framework to guide such efforts and support the development of adaptive and resilient communities.

One limitation of this study is the lack of access to actual traffic congestion data during the the Palisades and Eaton Fires. Without available open-source platforms to provide historical traffic information, we created a theoretical congestion model assuming vehicles moved at 10 kilometers per hour. While some commercial platforms offer historical traffic data, they require purchase and verification of data accuracy, which can be resource-intensive. Incorporating such data in future research could enhance the accuracy of accessibility measurements by reflecting actual travel conditions during the fire events.

This research assessed spatial accessibility on January 12, which marked the peak of wildfire activity. To gain a more comprehensive understanding of evacuation dynamics, future investigations should benefit from analyzing accessibility patterns throughout the entire duration of the wildfire incidents. This includes examining the initial days when fire activity was minimal, as well as the period following January 12 when containment efforts started working out. Additionally, future studies could explore shelter accessibility for vulnerable groups, such as older people and families without access to private vehicles, people under poverty level, communities with low household income, residential areas with low property values, and settlements in harsh environments to better inform equitable evacuation planning.

%% file: sections/append.tex
\section{Shelter Estimation}
\label{sec:shelters}

This section provides detailed descriptions of how shelter capacities were estimated for major emergency shelters in the study area. Estimations were based on publicly available building size data, architectural plans, and FEMA/Red Cross shelter guidelines. A common assumption was that approximately 30\% of total indoor space is unusable due to hallways, storage, restrooms, and other non-shelterable areas, and that each evacuee requires \SI{100}{ft^2} of space.

\begin{itemize}
    \item \textbf{Van Nuys/Sherman Oaks Recreation Center} \\
    Located at 14201 Huston St., Sherman Oaks, CA, the recreation center includes a gymnasium of approximately \SI{15000}{ft^2} \cite{paulmurdocharchitects}. Assuming 70\% of the space is usable, the effective area is \SI{10500}{ft^2}, yielding an estimated capacity of \textbf{105 people}.

    \item \textbf{Calvary Community Church} \\
    Situated at 5495 Via Rocas, Westlake Village, CA, the building covers approximately \SI{10573.27}{m^2}, as measured and verified using high-resolution satellite imagery. After excluding 30\% unusable space, the usable area equals \SI{7401.29}{m^2}. Converting to \SI{79720}{ft^2}, the estimated shelter capacity is \textbf{797 people}.

    \item \textbf{Pasadena Civic Auditorium} \\
    The Pasadena Convention Center (300 E Green Street, Pasadena, CA) provides roughly \SI{130000}{ft^2} of flexible indoor area \cite{visitpasadena}. Assuming 30\% unusable space, the usable area is \SI{91000}{ft^2}, supporting an estimated capacity of \textbf{910 people}.

    \item \textbf{Pomona Fairplex} \\
    The Fairplex complex offers about \SI{325000}{ft^2} of indoor exhibit space, including multiple halls suitable for emergency sheltering \cite{celeryoctopus}. Building 4 (\SI{105600}{ft^2}) and Buildings 5–8 (\SI{33600}{ft^2} each) plus Building 3 (\SI{11486}{ft^2}) together total \SI{150686}{ft^2}. Excluding 30\% unusable space leaves \SI{105480}{ft^2}, corresponding to an estimated capacity of \textbf{1,056 people}.

    \item \textbf{Glendale Civic Center} \\
    Located at 613~E~Broadway, Glendale, CA~91206, the building area, measured from high-resolution satellite imagery, is approximately \SI{28985}{ft^2} (\SI{2692.8}{m^2}). Considering that about 30\% of the space is typically unusable due to hallways, storage, and other non-shelterable areas, the estimated usable area is \SI{20290}{ft^2}. Assuming an average allocation of \SI{100}{ft^2} per person, the estimated shelter capacity of the facility is approximately \textbf{202 people}.
    
\end{itemize}

\begin{table}[h!]
\centering
\caption{Summary of estimated shelter capacities.}
\label{tab:shelter_capacity}
\resizebox{\textwidth}{!}{%
\begin{tabular}{lccc}
\toprule
\textbf{Shelter Name} & \textbf{Usable Area (ft$^2$)} & \textbf{Assumed Area per Person (ft$^2$)} & \textbf{Estimated Capacity} \\
\midrule
Van Nuys/Sherman Oaks Recreation Center & 10,500 & 100 & 105 \\
Calvary Community Church & 79,720 & 100 & 797 \\
Pasadena Civic Auditorium & 91,000 & 100 & 910 \\
Pomona Fairplex & 105,480 & 100 & 1,056 \\
Glendale Civic Center & 20,290 & 100 & 202 \\
\bottomrule
\end{tabular}%
}
\end{table}

\section{Road Network}
\label{sec:road_network}

\begin{figure}[tbp]
  \centering
  \includegraphics[width=0.95\linewidth]{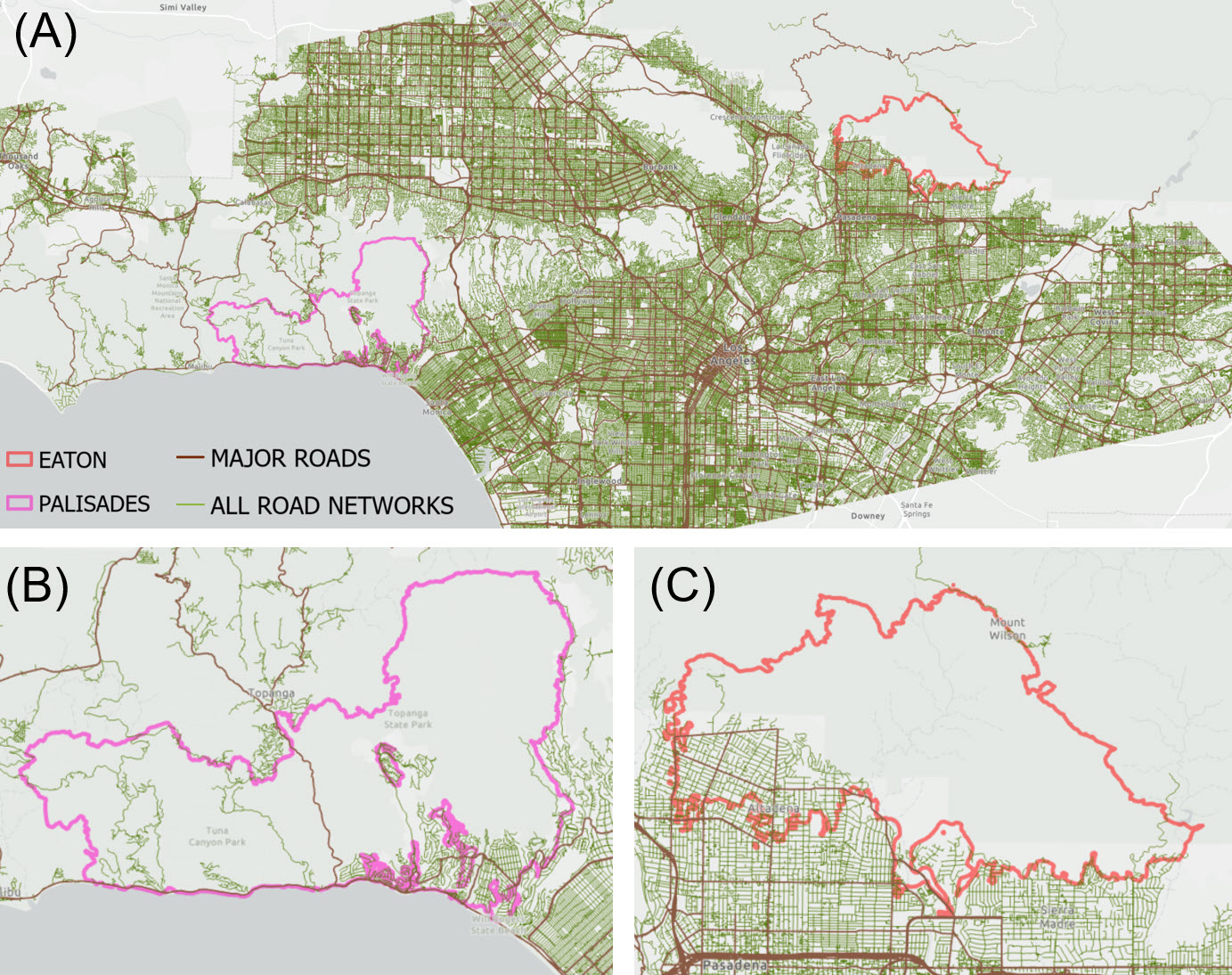}
  \caption{Road network data retrieved from OpenStreetMap (OSM) for the study area. (A) Full regional road network; (B) zoomed view of the Palisades Fire area; (C) zoomed view of the Eaton Fire area.}
  \label{fig:road_network}
\end{figure}

This section presents the visualization of the road network downloaded from OpenStreetMap (OSM) using the Python package \texttt{OSMnx}. Figure~\ref{fig:road_network}A shows both major and local roads across the entire study area, while Figures~\ref{fig:road_network}B and \ref{fig:road_network}C highlight the Palisades and Eaton Fire areas. The results demonstrate sparse road density within forested zones inside wildfire perimeters, whereas populations are largely concentrated in regions with denser road networks.

\section{Travel-Time Thresholds and Distance-Decay Specification}
\label{sec:accessibility}

In our spatial accessibility framework, we adopt a Gaussian distance-weighting function with a specified width ($\sigma = 30$ minutes) and impose a catchment travel-time threshold of 120 minutes (2 hours) under free-flow conditions. The Gaussian kernel is designed such that the weight $W_{ij}$ decreases sharply as travel time increases, approaching near-zero around 90 minutes. This formulation captures the rapid decline in accessibility as the effort required to reach a shelter increases. The 120-minute threshold serves as a hard cutoff under uncongested travel speeds, meaning that any origin located more than two hours from the nearest shelter is considered outside the practical catchment (accessibility = 0).

This modeling strategy aligns with established practices in generalized and enhanced Two-Step Floating Catchment Area (2SFCA) methods, which require defining an upper distance or time limit beyond which access is considered negligible \cite{luo2009enhanced, tao2020spatial}. Numerous studies have shown that accessibility results are highly sensitive to the choice of travel-time threshold. For example, \citep{he2021public} showed that reducing the maximum driving time from 60 to 30 minutes in Haiti decreased population coverage from 97.5\% to 85.4\%. Similarly, \citep{ermagun2019measuring,alam2021hurricane} evaluated multiple travel-time thresholds (15, 30, and 60 minutes) and found that accessibility declines substantially under more restrictive assumptions. Building on these studies, our use of a 90-minute effective decay and a 120-minute cutoff provides an upper-bound estimate that avoids prematurely excluding remote or traffic-congested areas from the analysis.

Empirical evidence supports the plausibility of this assumption. The NIST Camp Fire case study documented evacuees trapped in vehicles for over 90 minutes during the 2018 wildfire due to gridlock and limited egress routes \citep{nist2018campfire}. Similarly, FEMA’s all-hazard evacuation planning guidance recommends flexible clearance times that can exceed one to two hours depending on hazard type, population density, and network capacity \citep{fema1996slg101}. These findings emphasize that while actual travel during evacuations may far exceed two hours, using a 2-hour free-flow threshold provides a consistent and conservative upper bound for analytical purposes.
In summary, setting $\sigma = 30$ minutes and a 120-minute travel-time threshold captures the steep decline in accessibility associated with longer travel while maintaining computational tractability and empirical validity. The 90-minute point represents an effective limit of meaningful access, while the 120-minute cutoff ensures that accessibility beyond realistic uncongested travel conditions is excluded. This approach integrates real-world evacuation constraints with spatial accessibility theory to produce a robust and interpretable model.

\section{Assumptions for Evacuation Congestion and Travel-Speed Reduction}
\label{sec:workflow}

To conservatively represent near-source evacuation congestion under smoke-reduced visibility, we applied a 5~km buffer around evacuation and warning zones, within which vehicular speeds were capped at 10~km/h. This assumption is consistent with wildland--urban interface (WUI) traffic models that permit minimum speeds under dense smoke conditions \cite{intini2022wildfire, ronchi2023_verification_wui_evacuation}, empirical observations reporting evacuation corridor speeds of approximately 1--3~km/h during recent wildfire events \cite{winebusiness2025_sonoma_evacuations}, and planning-level analyses that adopt speed thresholds of approximately 5--15~mph (about 8--24~km/h) \cite{questhaven_evaplan_2021, SanDiego2022AnnexQ}. In addition, the selected 5~km buffer aligns with the 2--10~km spatial scales commonly used to delineate WUI evacuation buffer zones and study extents in prior research \cite{ZhaoWong2021BerkeleyTRR, Soga2024UCITS, GONG2024112179}.